\documentclass[12pt]{article}

\begin{document}

\title{The Theory of Stochastic Space-Time. II. Quantum Theory of
Relativity\footnote{Published in ''Z.Zakir (2003) \textit{Structure of
Space-Time and Matter}. CTPA, Tashkent''}}
\author{Zahid Zakir\thanks{E-Mail: zahid@in.edu.uz}\\Centre for Theoretical Physics and Astrophysics\\
P.O.Box 4412, Tashkent, 700000 Uzbekistan}
\date{January 4, 1999;\\
Revised: October 17, 2003.}

\maketitle
\begin{abstract}
The Nelson stochastic mechanics is derived as a consequence of the basic
physical principles such as the principle of relativity of observations and
the invariance of the action quantum. The unitary group of quantum mechanics
is represented as the transformations of the systems of perturbing devices. It
is argued that the physical spacetime has a stochastic nature, and that
quantum mechanics in Nelson's formulation correctly describes this stochasticity.
\end{abstract}

\textbf{1. Introduction.}

\textbf{2. Canonical transformations as transformations of the systems of
unperturbing devices.}

\textbf{3. Space and time in the systems of perturbing devices.}

\textbf{4. Relativity of perturbing observations and transformations of the
systems of perturbing devices.}

\textbf{5. The principle of constancy of action quantum.}

\textbf{6. Quantum principle of equivalence.}

\textbf{7}. \textbf{Conclusions.}

\section{Introduction}

A physical description is based on the analysis of results of observations,
however the laws of physics should not depend on methods of observations and a
choice of an observer. This condition we shall name as \textit{the principle
of relativity of observations, }and consider as a general principle to which
must satisfy all physical theories. In the paper we shall consider some
measurement procedures and mathematical structures, through which this general
principle is manifested in classical and quantum physics.

Particularly, the canonical transformations of Hamiltonian dynamics will be
treated as transformations of systems of unperturbing devices, the unitary
transformations of Hilbert space of states in quantum physics will be
represented as transformations of the systems of perturbing devices. This
means that one can take as the first principles of quantum mechanics the such
physical principles as the relativity under the systems of measuring devices
and the invariance of the fundamental constants $\hbar$ and $c$ \cite{ZZ1}
which lead to a stochastic geometry of the physical spacetime.

In the preceding paper \cite{ZZ2} some consequences of stochastic treatment of
gravitation has been considered. In this paper it will be shown that this
treatment can be derived from general physical principles.

\section{Canonical transformations as transformations of the systems of
unperturbing devices}

In the classical physics the system of coordinates have been constructed by
using the systems of devices allowing one to measure complete sets of
dynamical variables near each point of space. There it has been implicitly
supposed, that the basic equations of physics do not depend on a choice of
\textit{the} \textit{system of unperturbing devices}. We shall name this
statement as \textit{the} \textit{principle of relativity of unperturbing
observations }(PRUO).

If on a system of unperturbing devices an object is described by a full set of
dynamical variables - generalized coordinates and momentum $(q,p)$, and on
another system by another set of coordinates and momentum $(P,Q),$ so,
according to this principle, the equations of motion should not depend on the
transition from the first pair of variables to the second one. We see, that
the definition of transformations of the systems of unperturbing devices is
the same as the definition of the canonical transformations, and therefore, we
can suppose that \textit{the canonical invariance of the equations of motion
of classical physics is expression of the principle of relativity of
unperturbing observations.} This treatment allows one to describe the states
of mechanical objects by means of the systems of unperturbing devices in a
phase space with the symplectic structure, where the principle of relativity
of unperturbing observations is represented across the canonical group of symmetry.

From this point of view, \textit{the applicability of the Hamiltonian dynamics
}to various physical structures testifies that their dynamical variables can
be measured by the system of unperturbing devices, or they can be reduced to
the such variables. This interpretation partly explains the universality of
the Hamiltonian structures not only in the classical mechanics, but in other
fields of physics also.

\section{Spacetime in the systems of perturbing devices}

The existence of the Planck constant $\hbar$ as an action quantum requires to
extend the principle of relativity of observations to the \textit{systems of
perturbing devices} (SPD). Further we shall convince, that the such extension
is possible, and moreover, that the quantum theory can be considered as its
result. Here we shall consider the change of the structure of space and time
in SPD.

Let we have a system of perturbing devices as a set of classically described
measurement devices near each point of euclidean space. Let during the
measurements of coordinates and times of classical particles of small mass,
the particles scatter on devices at many points of their trajectories. Then,
in the limiting case, \textit{the trajectories of the such particles will be
similar to Brownian trajectories,} and the classical mechanics must be
replaced by the theory of Brownian motion. The system of perturbing devices
here plays a role of an environment with some diffusion coefficient and the
observables should be defined in a statistical ensemble of measurements.
Instead of the definite coordinates and momentum of the particles, here we
deal with the probability densities and the transition probability densities.

Two kinds of Brownian processes are known - the usual \textit{dissipative
(Wiener's) diffusion} and \textit{nondissipative (Nelson's) diffusion
}\cite{Ne}. The examples of the dissipative systems of perturbing devices with
Wiener's diffusion are the Wilson and bubble chambers in which a high energy
particle interacts with atoms of the medium along the trajectory and loses the energy.

At Nelson's diffusion the energy of an ensemble of particles is conserved and
\textit{the equations of diffusion are reversible in time} in contrary to
irreversible Wiener processes. Here we consider the construction of such
conservative systems of devices, in which the motion of particles represents
Nelson's diffusion \cite{ZZ1}. This is a set of large number massive screens
with \textit{infinite} number slits on each screen \cite{Fe}. Let the slits
have massive shutters, which must rapidly open and close the slits during very
short time. At the opening and closing of the shutters the sample particles
scatter on them. As the result, the energy and momentum of the particles may
change sufficiently, while the mean energy and momentum of the massive
shutters and screens do not change at the such scatterings. Therefore, the
tangent components of the momentum and kinetic energy of the scattering
particles change stochastically but their mean values remain unchanged.

Here a physical reason for the conservative behavior is the fact, that all
elements of the measuring devices are massive (macroscopic) objects,
essentially exceeding the masses of the sample particles. At the such ratio of
the masses, it is possible to consider the collisions of the classical
particles of small mass with very massive devices as absolutely elastic (on a
rest frame of the shutter). The energy conservation of the ensemble of
particles leads to the temporal reversibility of their equations of motions.

Thus in SPD the structure of spacetime becomes stochastic and the Galilean (or
Minkowski) geometry of spacetime should be replaced by the stochastic
geometry. In the dissipative SPD this is stochastic spacetime with Wiener's
measure, and in nondissipative SPD - spacetime with Nelson's measure.

\section{Relativity of perturbing observations and\textbf{\ }transformations
of SPD}

In classical mechanics with unperturbing devices it is insufficient how many
screens there are between two spatial points at particle's trajectory.
However, if we take into account the scattering of the particle on the screens
with moving shutters, then a number of scatterings becomes sufficient for the
final probability density of the such observed particles. At the increasing of
the number $N$ of the screens between initial and final positions, the
temporal intervals between the scatterings of the particles on neighbor
shutters decrease, and at $N\rightarrow\infty$ we have $\Delta t\rightarrow0$.
In fact, this is nothing but as some \textit{transformation of the system of
perturbing devices}, and the physically interesting variables are those which
tend to finite values at the such increasing of $N$.

Here we consider the such transformations of SPD and some conditions required
by the principle of relativity of observations. An ensemble of particles in
SPD is described by a probability density $\rho\left(  \mathbf{x},t\right)  $
and a transition probability density $p(\mathbf{x}^{\prime},t^{\prime
};\mathbf{x},t)$. In fact, there are two families of transition probabilities
$p_{\pm}$, where $p_{+}(\mathbf{x},t;\mathbf{x}_{0},t_{0})$, describes the
direct in time transition probabilities $t>t_{0}$, while $p_{-}(\mathbf{x}%
,t;\mathbf{x}_{0},t_{0})$ describes the backward in time transition
probabilities $t<t_{0}$. It is clear that in the conservative diffusion both
types must appear in the symmetric form. In the particular case of Nelson's
kinematics these two types of transition probabilities can be reduced to one a
function $S(\mathbf{x},t)$ and a diffusion coefficient $\nu$ \cite{Ne} (see
review \cite{Bl}).

A given SPD differs from another one only by the transition probability
densities $p_{\pm}(\mathbf{x},t;\mathbf{x}_{0},t_{0})$ describing the temporal
evolution of $\rho\left(  \mathbf{x},t\right)  $. For $\rho\left(
\mathbf{x},t\right)  $ on the first SPD one has:
\begin{equation}
\rho\left(  \mathbf{x},t\right)  =\int p_{\pm}(\mathbf{x},t;\mathbf{x}%
_{0},t_{0})\rho\left(  \mathbf{x}_{0},t_{0}\right)  d\mathbf{x}_{0},
\end{equation}
and on the second one with the same initial probability density $\rho
(\mathbf{x}_{0},t_{0})$:
\begin{equation}
\rho^{\prime}\left(  \mathbf{x},t\right)  =\int p_{\pm}^{\prime}%
(\mathbf{x},t;\mathbf{x}_{0},t_{0})\rho(\mathbf{x}_{0},t_{0})d\mathbf{x}_{0}.
\end{equation}

Here $\rho^{\prime}$ and $p_{\pm}^{\prime}$ transform as:%
\begin{equation}
\rho^{\prime}\left(  \mathbf{x},t\right)  =B(\mathbf{x},t)\rho\left(
\mathbf{x},t\right)  =(1+\delta B)\rho=\rho+\delta\rho,
\end{equation}
\begin{equation}
p_{\pm}^{\prime}(\mathbf{x},t;\mathbf{x}_{0},t_{0})=B(\mathbf{x},t)p_{\pm
}(\mathbf{x},t;\mathbf{x}_{0},t_{0})B^{-1}(\mathbf{x}_{0},t_{0})=
\end{equation}%
\begin{equation}
=p_{\pm}+(\delta Bp_{\pm}-p_{\pm}\delta B_{0})=p_{\pm}+\delta p_{\pm},
\end{equation}
where $B(\mathbf{x},t)$ is the operator for the transformations of
$\rho\left(  \mathbf{x},t\right)  $ at the changings of SPD.

The probability conservation conditions:
\begin{equation}
\int\rho\left(  \mathbf{x},t\right)  d\mathbf{x}=\int\rho^{\prime}\left(
\mathbf{x},t\right)  d\mathbf{x}=\int B(\mathbf{x},t)\rho\left(
\mathbf{x},t\right)  d\mathbf{x}=1,
\end{equation}
lead for small variations:%
\begin{equation}
\int\delta\rho\left(  \mathbf{x},t\right)  d\mathbf{x}=0,
\end{equation}
\begin{equation}
\int\delta B\left(  \mathbf{x},t\right)  \rho\left(  \mathbf{x},t\right)
d\mathbf{x}={\delta B}=0,
\end{equation}
since only the spatial distributions of $\rho\left(  \mathbf{x},t\right)  $
and $\delta B$ change at the such local deformations.

The velocities and diffusion coefficients of particles in SPD are defined as
the conditional expectations:%
\begin{equation}
\int d\mathbf{x}_{0}p_{\pm}(\mathbf{x},t;\mathbf{x}_{0},t_{0})(\mathbf{x-x}%
_{0})=\pm\mathbf{b}_{\pm}(\mathbf{x},t)\Delta t,
\end{equation}
\begin{equation}
\int d\mathbf{x}_{0}p_{\pm}(\mathbf{x},t;\mathbf{x}_{0},t_{0})(x_{i}%
-x_{i0})(x_{j}-x_{j0})=\pm2n_{\pm}(\mathbf{x},t)\delta_{ij}\Delta t,
\end{equation}
which transform at the SPD transformations as:%
\begin{equation}
\lim\limits_{\Delta t\rightarrow0}\int d\mathbf{x}_{0}B(\mathbf{x},t)p_{\pm
}(\mathbf{x},t;\mathbf{x}_{0},t_{0})B^{-1}(\mathbf{x}_{0},t_{0})(\mathbf{x}%
-\mathbf{x}_{0})=\pm\mathbf{b}_{\pm}^{\prime}(\mathbf{x},t)dt,
\end{equation}
\begin{equation}
\lim\limits_{\Delta t\rightarrow0}\int d\mathbf{x}_{0}B(\mathbf{x},t)p_{\pm
}(\mathbf{x},t;\mathbf{x}_{0},t_{0})B^{-1}(\mathbf{x}_{0},t_{0})(x_{i}%
-x_{i0})(x_{j}-x_{j0})=\pm2n_{ij\pm}^{\prime}(\mathbf{x},t)dt.
\end{equation}

\section{The principle of constancy of action quantum and the diffusion coefficient}

In classical mechanics it can be constructed SPD with arbitrary small
perturbations, and for a classical particle observing by SPD the conditional
expectation of the action function $\Delta A_{\pm}$ vanish for infinitesimal
temporal intervals:
\begin{equation}
\lim\limits_{\Delta t\rightarrow0}\int d\mathbf{x}_{0}p_{\pm}(\mathbf{x}%
,t;\mathbf{x}_{0},t_{0})\Delta A_{\pm}(\mathbf{x},t,\mathbf{x}_{0},t_{0})=0,
\end{equation}
which means that:
\begin{equation}
\lim\limits_{\Delta t\rightarrow0}\int d\mathbf{x}_{0}p_{\pm}(\mathbf{x}%
,t;\mathbf{x}_{0},t_{0})\left[  \frac{m\left(  \mathbf{x}-\mathbf{x}%
_{0}\right)  ^{2}}{\pm\Delta t}\right]  =0.
\end{equation}

Since the classical mechanics is not exact theory for the microscopic
phenomena, in the general case one must to take into account the existence of
the action quantum (Planck's constant) $\hbar$ which should
be\textit{\ invariant under the transformations of SPD}. The last statement we
shall call as \textit{the principle of constancy of action quantum }%
\cite{ZZ1}. Particularly, if the conditional expectation of $\Delta A_{\pm}$
is equal to $\hbar$ in one of SPD, then it should be equal to $\hbar$ in all
other SPD. So, we have the expression:
\begin{equation}
\lim\limits_{\Delta t\rightarrow0}\int d\mathbf{x}_{0}p_{\pm}(\mathbf{x}%
,t;\mathbf{x}_{0},t_{0})\left[  \frac{m\left(  \mathbf{x}-\mathbf{x}%
_{0}\right)  ^{2}}{\pm\Delta t}\right]  =\hbar,
\end{equation}
or in the SPD transformed form:
\begin{equation}
\lim\limits_{\Delta t\rightarrow0}\int d\mathbf{x}_{0}B(\mathbf{x},t)p_{\pm
}(\mathbf{x},t;\mathbf{x}_{0},t_{0})B^{-1}(\mathbf{x}_{0},t_{0})\left[
\frac{m\left(  \mathbf{x}-\mathbf{x}_{0}\right)  ^{2}}{\pm\Delta t}\right]
=\hbar.
\end{equation}
We see that, as the consequence of this principle, the conditional expectation
$E[dA_{\pm}\mid\mathbf{x}(t)]$ don't vanish at $\Delta t\rightarrow0.$ As the
result, we obtain the SPD covariant formula for the mean square values of
particle's coordinates:
\begin{equation}
\lim\limits_{\Delta t\rightarrow0}\int d\mathbf{x}_{0}p_{\pm}^{\prime
}(\mathbf{x},t;\mathbf{x}_{0},t_{0})\left(  \mathbf{x}-\mathbf{x}_{0}\right)
^{2}=\pm\frac{\hbar}{m}dt=\pm2\nu dt,
\end{equation}
where the diffusion coefficient $\nu=\hbar/2m$ is exactly the same as in
Nelson's stochastic formulation of quantum mechanics \cite{Ne}. But in the
stochastic mechanics the value of diffusion coefficient has been given equal
to $\nu=\hbar/2m$ by hand, whereas in the present treatment this formula
directly follows from the physically clear first principles.

Since for the SPD with the such invariant diffusion coefficients $n_{ij\pm
}^{\prime}=\pm2\nu\delta_{ij}$ the velocities $\mathbf{u}=(\mathbf{b}%
_{+}-\mathbf{b}_{-})/2$ can be expressed across $\rho\left(  \mathbf{x}%
,t\right)  $, the SPD transformations can be reduced to the transformations of
$\rho\left(  \mathbf{x},t\right)  $ and $\mathbf{v}=(\mathbf{b}_{+}%
+\mathbf{b}_{-})/2$ only:%
\begin{equation}
\delta\mathbf{u}=\nu\delta\left(  \frac{\mathbf{\nabla}\rho}{\rho}\right)  ,
\end{equation}
\begin{equation}
\delta\mathbf{v}=\frac{1}{m}\mathbf{\nabla}(\delta S),
\end{equation}
where $S(\mathbf{x},t)$ is some function, introduced instead of $\mathbf{v}%
(\mathbf{x},t)$ by the expression:
\begin{equation}
m\mathbf{v}=\mathbf{\nabla} S,
\end{equation}
and which can be derived from an initial Lagrangian after the special SPD
transformations \cite{Gu1}.

Finally, we have a functional ''phase space'' by the canonical pair $(\rho,S)$
for particle's motion in SPD, and the corresponding algebra of observables
\cite{Gu2} which are exactly equivalent to the Hilbert space of states and
operator algebra of the ordinary quantum mechanics.

\section{The quantum principle of equivalence}

The mass parameter $m$ in the diffusion coefficient of SPD $\nu_{d}%
=\hbar/2m_{in}$, considered in the previous Section, is the inertial mass
$m_{in}$ determined by the kinetic term of an action function describing the
scattering of particles on SPD.

In Nelson's stochastic mechanics we also have the diffusion coefficient with
the same inertial mass $\nu_{s}=\hbar/2m_{in}.$ The classical sample particle
with the inertial mass $m_{in}$ freely moves in the stochastic space, and the
process represents the conservative diffusion on some background.

In quantum mechanics, rewritten in terms of the stochastic processes, we have
the effective diffusion coefficient $\nu_{q}=\hbar/2m_{q}$, where $m_{q}$ is
some mass parameter, determining the fluctuations of coordinates of quantum
particles in flat and regular Galilean space and time \cite{Sm}. As it was
shown from analysis of the Lamb shift data, the quantum diffusion mass $m_{q}$
is equal to the inertial mass with high accuracy \cite{Sm}:
\begin{equation}
(m_{in}-m_{q})/m_{in}\leq10^{-13}.
\end{equation}

Therefore, two kind of masses of the particle are equal to each other:
\begin{equation}
m_{in}=m_{q},
\end{equation}
and this fact is very important for the understanding of a geometrical nature
of quantum phenomena.

Firstly this means the equivalence of the motion of the classical particle on
SPD in the ordinary smooth space to the motion of the classical particle in
the stochastic space with unperturbing devices. This fact means also the
equivalence of the transition to SPD and the quantum mechanical description
(quantization) of the motion of the classical particle.

We can demonstrate this situation in the simple double slit experiment. Let we
have a source and a detector of the particles and there exists between them a
screen with two slits. Let the particles, emitted by the source and penetrated
across the slits, have been registered by the detectors. After the repeating
the experiment many times, an observer obtains the interference picture on the
detectors. The observer, which have only photographic plate with interference
picture, can not distinguish differences between three interpretations:

a) Space is empty and \textit{euclidean}, but \textit{the particles have
''quantum'' properties} (wave function) leading to the interference. This is
the \textit{quantum mechanical} interpretation;

b) Space is empty and \textit{stochastic}, and the motion of the classical
particles\textit{\ }on this background leads to the interference. This is the
interpretation of the stochastic mechanics;

c) Space is smooth and euclidean, but \textit{it not empty,} and there exist
infinity number devices near each point of space. The classical particles
interact with this medium of devices, and as the result, the observer detects
the interference. This treatment based on the thought experiment illustrating
the principle of relativity of observations.

Further we will call this fact as \textit{the quantum principle of equivalence
} \cite{ZZ1}. Can we conclude from these statements that quantum mechanics is
the stochastic geometry of spacetime and that the stochastic mechanics is a
true physical formulation of quantum mechanics? \textit{Is the physical
spacetime stochastic?}

For the answer to these questions, let us remind an analogue with Einstein's
proof of a geometrical nature of gravitation. From the equality of the
inertial and gravitational masses in general relativity, one has
indistinguishability for the observer of three treatments in the explaining of
acceleration of a sample particle:

a) spacetime is euclidean, the reference frame is inertial, but there
\textit{exist a gravitational field} with the potential $\varphi$ (the field treatment);

b) \textit{spacetime is Riemannian}, the reference frame is (locally)
inertial, and no any external field (geometric treatment);

c) spacetime is euclidean in a global inertial frame, but the
\textit{reference frame of the observer is accelerated}, and no external field
(kinematic treatment).

After analyzing these situations in general relativity, the geometric theory
of gravitation has been established. Analogously, we can conclude, that the
quantum principle of equivalence allows us to justify the stochastic
geometrical version of quantum mechanics. It is important, that quantum
mechanics is nothing but as the stochastic geometry of spacetime, has the very
important consequence as an explanation of \textit{gravitation as
inhomogeneous quantum diffusion} \cite{ZZ2}.

\end{document}